\documentclass[letterpaper,12pt]{article}
\usepackage{authblk}
\usepackage[utf8]{inputenc}
\usepackage[english]{babel}
\usepackage{amsmath}
\usepackage{amssymb}
\usepackage{graphicx}
\usepackage{multirow}
\usepackage{epsfig}
\usepackage{rotating}
\usepackage{longtable} 
\usepackage{caption}
\usepackage{subcaption}
\usepackage{url}
\usepackage{pdflscape}
\usepackage{lscape}
\usepackage{cancel}
\usepackage{setspace}
\pdfoutput=1
\usepackage{csquotes}
\usepackage{placeins}
\usepackage[left=2cm,top=3cm,right=2cm,bottom=3cm]{geometry}
\usepackage{tabularx}
\usepackage{adjustbox}
\usepackage{color,xcolor}

\usepackage{hyperref}

\usepackage{mathtools}
\usepackage{latexsym}

\title{One texture zero for Dirac neutrinos in a diagonal charged lepton basis}

\author[a]{Richard H. Benavides}
\author[b]{Yessica Lenis}
\author[a]{John D. Gómez}
\author[b]{William A. Ponce}

\affil[a]{Facultad de Ciencias Exactas y Aplicadas, Instituto Tecnol\'{o}gico Metropolitano, Calle 73 N° 76-354 via el volador, Medell\'{i}n, Colombia.} 
\affil[b]{Instituto de Física, Universidad de Antioquia, A.A. 1226, Medellín, Colombia.}

\begin{document}

\maketitle
\abstract{An analytic and numerical systematic study of the neutrino mass matrix with one texture zero is presented in a basis where the charged leptons are diagonal. Under the assumption that neutrinos are Dirac particles, the analysis is carried out in detail for the normal and inverted hierarchy mass spectrum. Our study is performed without any approximations, first analytically and then numerically, using current neutrino oscillation data. The analysis constrains the parameter space in such a way that, among the six possible one-texture-zero patterns, only four are favored in the normal hierarchy and, one in the inverted hierarchy, by current oscillation data at the $3 \sigma$ level. Phenomenological implications for the lepton CP-violating phase and neutrino masses are also explored.
}

\section{Introducción}
Although the gauge boson sector of the Standard Model (SM) with the $SU(3)_c\otimes SU(2)_L\otimes U(1)_Y$ local gauge symmetry has been very successful so far (with $SU(3)_c$ confined and $SU(2)_L$ spontaneously broken via the Higgs mechanism~\cite{dms}), its Yukawa sector is still poorly understood. Questions related to this sector such as the number of families in nature, the hierarchy of the charged fermion mass spectrum, the smallness of the neutrino masses, the quark mixing angles, the neutrino oscillation parameters, and the origin of CP violation remains as open questions to date.
Besides, in the context of the SM there is a lack of explanation for the dark matter and the dark energy observed at present in the universe.

In the context of the SM a neutrino flavor created by the weak interaction and associated with a charged lepton will maintain its flavor, which implies that lepton flavor is conserved and neutrinos are massless. Moreover, from oscillation experiments we know that neutrinos are massive particles and that they oscillate from one flavor to another, with the results sensitive only to the squared mass difference. We still do not know the mass of any of the light neutrinos. However, from cosmology, we know  an upper
limit for the sum of the three light neutrino masses: $\sum m_\nu < 0.12\ \text{eV} \quad \text{(95\% CL)}$ 
\cite{Planck2018,eBOSS2021,DES2022}, while from tritium beta decay there is room for an effective mass of the electron neutrino $\langle m_{\beta} \rangle < 0.8\ \text{eV} \quad \text{(90\% CL)}$ \cite{KATRIN2022,KATRIN2021,Drexlin2013}.

Current neutrino experiments are measuring the neutrino mixing parameters with unprecedented accuracy.
The next generation of neutrino experiments will be sensitive to subdominant neutrino oscillation effects that can,
in principle, provide information on the yet unknown
neutrino parameters: the Dirac CP-violating phase in the
Pontecorvo-Maki-Nakagawa-Sakata (PMNS) mixing
matrix $U_{PMNS}$, the neutrino mass ordering, and the octant
of the mixing angles.

To date, the solar and atmospheric neutrino oscillations have established the following values to 3 sigma of
the deviation to normal ordering: \cite{nufit}
\begin{eqnarray*}
\Delta m^2_{Atm}&=&(2.46-2.61)\times 10^{-3} eV^2=\Delta m^2_{32},\\
\Delta m^2_{Sol} &=& (6.92 - 8.05)\times 10^{-5} eV^2 = \Delta m^2_{21} ,\\
\sin^2 \theta_{Atm} &=& (4.30 - 6.96) \times 10^{-1} = \sin^2 \theta_{23} ,\\
\sin^2 \theta_{Sol} &=& (2.75-3.45) \times 10^{-1}  = \sin^2 \theta_{21} , \\
\sin^2 \theta_{Reac} &=& (2.02 - 2.38)\times 10^{-2}  = \sin^2 \theta_{13}.
\end{eqnarray*}

Numbers obtained under the assumption that for the charged lepton sector the weak basis and the flavor basis are the same.

In this work we present a systematic  study of the neutrino mass matrix $M_\nu$ with one texture zero, in a basis where the three charged leptons are diagonal, as an extension of the already presented study of the case with two texture zeros~\cite{Lenis:2023lgq,Lenis:2023wyr}.

In our analysis, for each one of the six different one texture zero in $M_\nu$, we carry, first an analytic, and then a statistical fit of the oscillation angles, in order to limit the parameter space, getting in this way neat predictions for the neutrino masses.

\section{The Model}
The model in which we are going to work is a simple extension of the SM, with the following three new ingredients added:
\begin{enumerate}
 \item We extend the electroweak sector of the standard model with three right handed neutrinos,  
 $(\nu_{\alpha R};\;\alpha = e,\mu,\tau)$.
 \item The charged lepton mass matrix is diagonal in the weak basis.
 \item Majorana masses are forbidden.
\end{enumerate}
\subsection{Neutrino mass matrix}
According to the previous hypothesis, for the charged lepton sector we have
\begin{equation}\label{mslcar}
M_l=\left(\begin{array}{ccc}m_e & 0 & 0 \\ 0 & m_\mu & 0 \\ 0 & 0 & m_\tau 
\end{array}\right),
\end{equation}

which implies that the most general $ 3\times 3$ neutrino mass matrix is complex, however, a complex matrix can always be decompose as the product of a Hermitian and a unitary matrix, according to the polar decomposition theorem $(M=HU)$,  where the unitary factor can be absorbed into the $SU(2)_L$ singlet gauge structure\cite{Prasolov1994,HornJohnsonMatrixAnalysis,Bhatia1997}. Then we can assume it as Hermitian matrix without loss of generality, that means,

\begin{eqnarray}\nonumber
 M_\nu&=&\left(\begin{array}{ccc} m_{\nu_e\nu_e} & m_{\nu_e\nu_\mu} & m_{\nu_e\nu_\tau} 
\\ m_{\nu_\mu\nu_e} & m_{\nu_\mu\nu_\mu} & m_{\nu_\mu\nu_\tau} 
\\ m_{\nu_\tau\nu_e} & m_{\nu_\tau\nu_\mu} & m_{\nu_\tau\nu_\tau} 
\end{array}\right)= 
 V_{PMNS}\left(\begin{array}{ccc} m_1 & 0 & 0 \\ 0 & m_2 & 0 \\ 0 & 0 & m_3 
\end{array}\right)V_{PMNS}^\dagger \\ \label{mnud}
&=&{\begin{bmatrix}U_{e1}&U_{e2}&U_{e3}\\U_{\mu 1}&U_{\mu 2}&U_{\mu 3}\\
U_{\tau 1}&U_{\tau 2}&U_{\tau 3}\end{bmatrix}}
{\begin{bmatrix}m_1 & 0 & 0 \\ 0 & m_2 & 0 \\ 0 & 0 & m_3 \end{bmatrix}}
{\begin{bmatrix}U_{e1}^*&U_{\mu 1}^*&U_{\tau 1}^*\\U_{e2}^*&U_{\mu 2}^*&U_{\tau 2}^*\\
U_{e3}^*&U_{\mu 3}^*&U_{\tau 3}^*\end{bmatrix}}\\ \nonumber
&=&{\begin{bmatrix}U_{e1}&U_{e2}&U_{e3}\\U_{\mu 1}&U_{\mu 2}&U_{\mu 3}\\
U_{\tau 1}&U_{\tau 2}&U_{\tau 3}\end{bmatrix}}
{\begin{bmatrix}m_1U_{e1}^*&m_1U_{\mu 1}^*&m_1U_{\tau 1}^*\\
m_2U_{e2}^*&m_2U_{\mu 2}^*&m_2U_{\tau 2}^*\\
m_3U_{e3}^*&m_3U_{\mu 3}^*&m_3U_{\tau 3}^*\end{bmatrix}}
\end{eqnarray}
where the mixing matrix $V_{PMNS}$ for Dirac neutrinos is parametrized in 
the usual way as:
\begin{equation}
{\displaystyle{\begin{aligned}&{\begin{bmatrix}1&0&0\\
0&c_{23}&s_{23}\\0&-s_{23}&c_{23}\end{bmatrix}}
{\begin{bmatrix}c_{13}&0&s_{13}e^{-i\delta _{\text{CP}}}\\
0&1&0\\-s_{13}e^{i\delta _{\text{CP}}}&0&c_{13}\end{bmatrix}}
{\begin{bmatrix}c_{12}&s_{12}&0\\-s_{12}&c_{12}&0\\0&0&1\end{bmatrix}}\\
&={\begin{bmatrix}c_{12}c_{13}&s_{12}c_{13}&s_{13}e^{-i\delta _{\text{CP}}}\\
-s_{12}c_{23}-c_{12}s_{23}s_{13}e^{i\delta _{\text{CP}}}&
c_{12}c_{23}-s_{12}s_{23}s_{13}e^{i\delta _{\text{CP}}}&
s_{23}c_{13}\\s_{12}s_{23}-c_{12}c_{23}s_{13}e^{i\delta _{\text{CP}}}&
-c_{12}s_{23}-s_{12}c_{23}s_{13}e^{i\delta _{\text{CP}}}&c_{23}c_{13}\end{bmatrix}};
\end{aligned}}} 
\end{equation}
where $Diag(m_1,m_2,m_3)$ refers to the neutrino mass eigenvalues, and 
$c_{ij}=\cos\theta_{ij}$ and $s_{ij}=\sin\theta_{ij}$ are the cosine and sine of the oscillation angles $\theta_{ij},\;\;i < j=1,2,3$ in the standard parametrization.

Now, due to the hermiticity constraint, the elements of $M_\nu$ satisfy:
$m_{\nu_e\nu_e}=m_{\nu_e\nu_e}^*,\; 
m_{\nu_\mu\nu_\mu}=m_{\nu_\mu\nu_\mu}^*,\; 
m_{\nu_\tau\nu_\tau}=m_{\nu_\tau\nu_\tau}^*$, and 
$m_{\nu_\mu\nu_e}=m_{\nu_e\nu_\mu}^*$, 
$m_{\nu_\tau\nu_e}=m_{\nu_e\nu_\tau}^*$ and 
$m_{\nu_\mu\nu_\tau}=m_{\nu_\tau\nu_\mu}^*$.

In our analysis we use the numerical values for the entries of the $U_{PMNS}$ measured at $3\sigma$ ranges presented in the literature~\cite{JOUR}, and quoted above.


\subsection{Counting parameters}
When the mass matrices for the lepton sector are given by 
(\ref{mslcar}) and (\ref{mnud}), we have that in the weak basis, the most general hermitian mass matrix $M_\nu$ has six real parameters and three phases that we can use to 
explain seven physical parameters: three real oscillating angles 
$\theta_{12},\;\;\theta_{13}$ and $\theta_{23}$, three real neutrino masses $m_1,\; m_2$ and $m_3$, and one CP violating phase $\delta$. So, in principle, we have a redundant number of parameters (two more phases).

Now, contrary to the quark sector, we can not introduce texture zeros via weak basis transformations \cite{Branco1989,Branco2000,Fritzsch2000,Ponce:2011qp,Ponce:2013nsa} in the mass matrix $M_\nu$ due to the fact that it will change the charged lepton diagonal mass matrix. But as it is shown elsewhere \cite{Lenis:2023lgq,Lenis:2023wyr,Benavides:2020pjx}, the ``Weak basis transformations'' can be used to eliminate the two redundant phases, ending in our study with just seven parameters, six real and one phase, enough to accommodate in principle the seven physical parameters. 

Now, since the number of analytic parameters is equal to the number of physical ones, one (or more) texture zeros in the mass matrix $M_\nu$  will imply relationships between the physical parameters, in particular, for one texture zero we may expect one relationship between the neutrino masses and the oscillating  angles in the $U_{PMNS}$ matrix (as it happens for example in the quark sector \cite{Ponce:2011qp,Ponce:2013nsa}).\\

The origin of texture zeros has been extensively discussed in the literature (see, for instance, \cite{Frampton2002,FritzschXing2011,Dev2015,Mondragon1999,Mondragon2007}). Such vanishing entries can indeed arise from the imposition of discrete flavor symmetries or specific group structures acting on the lepton fields.

\section{One texture zero}\label{subsec:fermionico}
The introduction of texture zeros in a general mass matrix has been an outstanding hypothesis that may provide relationships in the lepton sector, between the oscillation angles and the mass eigenvalues.

As discussed above, the six real mathematical parameters of the most general Hermitian mass matrix for Dirac neutrinos provide  sufficient room to accommodate the five real experimental values with no prediction at all. One texture zero, even though should not conduce to any prediction either, limits the parameter space enough to accommodate or not the experimental measured numbers.

In the following, for the case of ``Normal Ordering" (NO) and "Inverted ordering" (IO), we will study the six different cases of one texture zero in the hermitian Dirac mass matrix $M_\nu$, looking in special for the lightest Dirac neutrino mass, which consequently implies the knowledge of the neutrino mass spectrum.

Our aim in this study is to carry, first an analytic, and then  an statistical analysis of the parameter space, when one texture zero is introduced in $M_\nu$.

For the implications of texture zeros in $M_\nu$, two cases must be analyzed: texture zeros in the diagonal, and texture zeros outside the diagonal.

\subsubsection{Diagonal texture zeros}
Let us start assuming that $m_{\nu_e\nu_e}=0$ and see its implications: 

From equation (\ref{mnud}) we have: 
\begin{equation}\label{tz12d}
 m_{\nu_e\nu_e}=m_1|U_{e1}|^2 +m_2|U_{e2}|^2 +m_3|U_{e3}|^2 =0,
\end{equation}
dividing by $m_3$ and using the unitary constraint of matrix  $U$; that is 
$|U_{e1}|^2 +|U_{e2}|^2 +|U_{e3}|^2=1$ we can writte (\ref{tz12d}) as: 
\[\frac{m_1}{m_3}|U_{e1}|^2+\frac{m_2}{m_3}|U_{e2}|^2+1-|U_{e1}|^2-|U_{e2}|^2=0;\]
which we can rearrenge as:
\begin{equation}\label{mnudp1}
 |U_{e2}|^2=\frac{m_3}{m_3-m_2}-\frac{m_3-m_1}{m_3-m_2}|U_{e1}|^2.
\end{equation}

By putting this equation~(\ref{mnudp1}) in terms of the physical parameters we obtain:
\begin{equation}\label{one}
s_{12}^2c_{13}^2=\frac{m_3}{(m_3-m_2)}-\frac{(m_3-m_1)}{(m_3-m_2)}c_{12}^2c_{13}^2,
\end{equation}
As anticipated above, producing a relationship between the neutrino masses and the oscillation parameters. 

In a similar way for $m_{\nu_\mu\nu_\mu}=0$ we have:
\begin{equation}\label{mnudp2}
 |U_{\mu 2}|^2=\frac{m_3}{m_3-m_2}-\frac{m_3-m_1}{m_3-m_2}|U_{\mu 1}|^2,
\end{equation}
and for $m_{\nu_\tau\nu_\tau}=0$ we have:
\begin{equation}\label{mnudp2}
 |U_{\tau 2}|^2=\frac{m_3}{m_3-m_2}-\frac{m_3-m_1}{m_3-m_2}|U_{\tau 1}|^2.
\end{equation}

The former three cases can be summarized as:
\begin{equation}\label{mnudpg}
 |U_{\alpha 2}|^2=\frac{m_3}{m_3-m_2}-\frac{m_3-m_1}{m_3-m_2}|U_{\alpha 1}|^2, 
\end{equation}
for $\alpha=e$ if $m_{\nu_e\nu_e}=0$; $\alpha=\mu$ if $m_{\nu_\mu\nu_\mu}=0$, 
and $\alpha=\tau$ if $m_{\nu_\tau\nu_\tau}=0$.

\subsubsection{Texture zeros outside the diagonal}
Let us consider now a texture zero outside the diagonal. Let us start with 
$m_{\nu_e\nu_\mu}=0$ (notice that $m_{\nu_\mu\nu_e}=m_{\nu_e\nu_\mu}^*=0$).

For this situation equation (\ref{mnud}) implies that: 
\begin{equation}\label{tz12nd}
 m_{\nu_e\nu_\mu}=m_1U_{e1}U_{\mu 1}^* +m_2U_{e2}U_{\mu 2}^* +m_3U_{e3}U_{\mu 3}^* =0,
\end{equation}
which dividing by $m_3$ and using the orthogonality condition 
$U_{e1}U_{\mu 1}^* +U_{e2}U_{\mu 2}^* +U_{e3}U_{\mu 3}^*=0$ can be written as: 

\begin{equation}\label{tz12p}
 \left(\frac{m_1}{m_3}-1\right)U_{e1}U_{\mu 1}^* +
\left(\frac{m_2}{m_3}-1\right)U_{e2}U_{\mu 2}^*=0,
\end{equation}
multiplying by $U_{e2}^*U_{\mu 2}$ and rearrenging we have 
\begin{equation}\label{tz12pp}
 \left(\frac{m_1}{m_3}-1\right)U_{e1}U_{\mu 1}^*U_{e2}^*U_{\mu 2} +
\left(\frac{m_2}{m_3}-1\right)|U_{e2}|^2|U_{\mu 2}|^2=0, 
\end{equation}
which we can finally writte as
\begin{equation}\label{tz12ppp}
U_{e1}U_{\mu 1}^*U_{e2}^*U_{\mu 2} +
\left(\frac{m_3-m_2}{m_3-m_1}\right)|U_{e2}|^2|U_{\mu 2}|^2=0, 
\end{equation}
equation, that toghether with its complex conjugate, can be separated in two parts: 
a real part equal to zero and an imaginary part also equal to zero (notice that 
for an hermitian matrix its eigenvalues must be real but not necesarily possitive).

As $m_{\nu_\mu\nu_e}$ must also be equal to zero, the two relations must also be 
equivalent to take the real part and the imaginary part in (\ref{tz12ppp})
equal to zero. As the second term in (\ref{tz12ppp}) is real, to take the 
imaginary part equal to zero produces 
\begin{equation}
 Im(U_{e1}U_{\mu 1}^*U_{e2}^*U_{\mu 2})=J=0; 
\end{equation}
which means that this texture zero is associated to a Jarlskog invariant 
equal to zero and no CP violation is pressent for this texture zero.

In a similar way for $m_{\nu_e\nu_\tau}=0$, we have: 
\begin{equation}\label{tz13nd}
 m_{\nu_e\nu_\tau}=m_1U_{e1}U_{\tau 1}^* +m_2U_{e2}U_{\tau 2}^* + 
 m_3U_{e3}U_{\tau 3}^* =0.
\end{equation}
Dividing by $m_3$ and using the orthogonality relationship 
$U_{e1}U_{\tau 1}^* +U_{e2}U_{\tau 2}^* +U_{e3}U_{\tau 3}^*=0$ we can writte  
(\ref{tz13nd}) como: 
\begin{equation}\label{tz13p}
 \left(\frac{m_1}{m_3}-1\right)U_{e1}U_{\tau 1}^* +
\left(\frac{m_2}{m_3}-1\right)U_{e2}U_{\tau 2}^*=0,
\end{equation}
multiplying by $U_{e2}^*U_{\tau 2}$ and reordering, we have 
\begin{equation}\label{tz13pp}
 \left(\frac{m_1}{m_3}-1\right)U_{e1}U_{\tau 1}^*U_{e2}^*U_{\tau 2} +
\left(\frac{m_2}{m_3}-1\right)|U_{e2}|^2|U_{\tau 2}|^2=0, 
\end{equation}
which in turns implies 
\begin{equation}\label{tz13ppp}
U_{e1}U_{\tau 1}^*U_{e2}^*U_{\tau 2} +
\left(\frac{m_3-m_2}{m_3-m_1}\right)|U_{e2}|^2|U_{\tau 2}|^2=0, 
\end{equation}
which again produces  
\begin{equation}
 Im\,(U_{e1}U_{\tau 1}^*U_{e2}^*U_{\tau 2})=J=0. 
\end{equation}
The former means that this texture zero outside the diagonal, and also the 
former case, are associated to an Jarlskog invariant equal to zero and 
again, there is no CP violation for this case.

In a similar way we have for $m_{\nu_\mu\nu_\tau}=0$ that 
\begin{equation}\label{tz23nd}
 m_{\nu_\mu\nu_\tau}=m_1U_{\mu 1}U_{\tau 1}^* +m_2U_{\mu 2}U_{\tau 2}^* + 
 m_3U_{\mu 3}U_{\tau 3}^* =0,
\end{equation}
which divided by $m_3$ and making use of the appropriate orthogonality relationship, we have
\begin{equation}\label{tz23p}
 \left(\frac{m_1}{m_3}-1\right)U_{\mu 1}U_{\tau 1}^* +
\left(\frac{m_2}{m_3}-1\right)U_{\mu 2}U_{\tau 2}^*=0,
\end{equation}
which multiplied by $U_{\mu 2}^*U_{\tau 2}$ we get  
\begin{equation}\label{tz23ppp}
U_{\mu 1}U_{\tau 1}^*U_{\mu 2}^*U_{\tau 2} +
\left(\frac{m_3-m_2}{m_3-m_1}\right)|U_{\mu 2}|^2|U_{\tau 2}|^2=0, 
\end{equation}
which again takes as to 
\begin{equation}
 Im\,(U_{\mu 1}U_{\tau 1}^*U_{\mu 2}^*U_{\tau 2})=J=0; 
\end{equation}

\section{Numerical results}\label{sec:4}

After the spontaneous symmetry breaking of the local gauge symmetry, the Lagrangian mass term for the lepton sector is given by

\begin{equation}
 -\mathcal{L}=\bar{\nu}_LM_\nu \nu_R+\bar{l}_LM_ll_R + h.c.
\end{equation}
where $M_\nu$ and $M_l$ are, in general, complex matrices. We are going to analyze each one of the texture forms with one texture zero in the neutral sector and choose the charged lepton sector in diagonal form.

\subsection{Textures Forms}\label{sec:4.1}

There are six one texture zero different patterns in $M_\nu$ for which we use the following notation:

\[
T_1 =
\begin{pmatrix}
0 & a & b \\
a^* & c & d \\
b^* & d^* & e
\end{pmatrix}, \quad
T_2 =
\begin{pmatrix}
a & b & c \\
b^* & 0 & d \\
c^* & d^* & e
\end{pmatrix}, \quad
T_3 =
\begin{pmatrix}
a & b & c \\
b^* & d & e \\
c^* & e^* & 0
\end{pmatrix},
\]

\[
T_4 =
\begin{pmatrix}
a & 0 & b \\
0 & c & d \\
b^* & d^* & e
\end{pmatrix}, \quad
T_5 =
\begin{pmatrix}
a & b & 0 \\
b^* & c & d \\
0 & d^* & e
\end{pmatrix}, \quad
T_6 =
\begin{pmatrix}
a & b & c \\
b^* & d & 0 \\
c^* & 0 & e
\end{pmatrix}.
\]

In the following, we will perform a numerical analysis for each of the textures.

\subsection{Texture \texorpdfstring{$T_1$}{T1}}\label{sec:4.2}

The texture $T_1$, featuring a vanishing $(1,1)$ element, provides a good fit to the observed neutrino oscillation data. In this case, the parameters $a$, $b$, and $d$ are complex, while $c$ and $e$ are real. The mass matrix \( M_\nu \) must satisfy the diagonalization condition:

\[
U^\dagger M_\nu U = \text{diag}(m_1, m_2, m_3).
\]

Where $U$ is the same $U_{PMNS}$  matrix known experimentally.  To quantify the deviation between the observables predicted by the mass matrix \( M_\nu \) and the experimentally measured values, we construct a chi-squared error function:

\[
\chi^2 = \sum_i \left( \frac{x_i^{\text{pred}} - x_i^{\text{obs}}}{\sigma_i} \right)^2,
\]
where \( x_i \) represents the relevant physical observables (such as oscillation angles and mass-squared differences), \( x_i^{\text{pred}} \) are the theoretical predictions obtained from the diagonalization of \( M_\nu \), \( x_i^{\text{obs}} \) are the experimentally measured values, and \( \sigma_i \) are the corresponding experimental uncertainties, as provided in the introduction.

The goal is to find the set of parameters in \( M_\nu \) that minimizes \( \chi^2 \), ensuring that the matrix reproduces the observed neutrino oscillation data within the allowed confidence level. The same procedure is applied to the other texture types.

\subsubsection{Mass Matrix Parameters}
Our numerical analysis shows that the mass matrix elements are:
\[
M_\nu = \begin{pmatrix}
0 & 0.00516 + 0.0057i & -0.00411 + 0.00461i \\
0.00516 - 0.0057i & 0.0281 & 0.02225 - 0.00005i \\
-0.00411 - 0.00461i & 0.02225 + 0.00005i & 0.02507
\end{pmatrix}~\text{eV}
\]


with the resulting eigenvalues:
\[
\begin{aligned}
m_1 = -0.00542~\text{eV}, \,\,\,
m_2 = 0.00860~\text{eV}, \,\,\,
m_3 = 0.04999~\text{eV}, \,\,\,
\sum_{i=1}^3 |m_i| = 0.06402~\text{eV}.
\end{aligned}
\]

This spectrum is consistent with a normal mass ordering and complies with cosmological bounds on the sum of neutrino masses \cite{planck2020}, and 

\[
\begin{aligned}
\Delta m_{21}^2 = 4.46 \times 10^{-5}~\text{eV}^2, \,\,\,
\Delta m_{31}^2 = 2.50 \times 10^{-3}~\text{eV}^2,
\end{aligned}
\]
values which are in excellent agreement with experimental constraints from solar and atmospheric neutrino data \cite{Fukuda:1998mi}.
These values are evaluated with the best fit point for the minimum $\chi^2=1.43\times10^{-3}$.


The former values allow us to obtain the following mixing matrix: 

\[
U_{PMNS} = \begin{pmatrix}
 -0.818602 + 0.0905867 i & -0.537971 - 0.101074 i & -0.0162952 - 0.147616 i \\
 0.374339 - 0.153715 i & -0.541351 - 0.128917 i & -0.725065 - 0.0290143 i \\
 -0.3974 & 0.625053 & -0.67185
\end{pmatrix}.
\]

Matrix which is unitary and consistent with the standard parameterization of the lepton mixing matrix. The oscillation angles  extracted from this matrix are:

\[
\begin{aligned}
\theta_{13} = \arcsin\left( |U_{13}| \right), \,\,\,
\theta_{12} = \arctan\left( \frac{|U_{12}|}{|U_{11}|} \right), \,\,\,
\theta_{23} = \arctan\left( \frac{|U_{2 3}|}{|U_{3 3}|} \right).
\end{aligned}
\]

And to extract the Dirac CP-violating phase \( \delta \), one can use the Jarlskog invariant \( J \), defined as:

\[
J = \text{Im}\left( U_{11} U_{22} U_{12}^* U_{21}^* \right)
= \frac{1}{8} \sin 2\theta_{12} \sin 2\theta_{23} \sin 2\theta_{13} \cos\theta_{13} \sin\delta.
\]

\[
\begin{aligned}
\theta_{12} = 33.61^\circ, \,\,\,
\theta_{13} = 8.54^\circ, \,\,\,
\theta_{23} = 47.21^\circ, \,\,\,
\delta = 270.02^\circ.\,\,\,
\end{aligned}
\]
\noindent 
Numbers in quite good agreement with the experimental meassured values at the $3\sigma$ level reported by the NuFIT collaboration~\cite{JOUR}.\\

This texture $T_1$ is highly successful in accommodating the current experimental data on neutrino mixing and mass-squared differences, with a physically acceptable prediction for the CP-violating phase. The choice of a zero value at the $(1,1)$ position can be motivated by underlying flavor symmetries such as $L_e - L_\mu - L_\tau$, and encourage us to further studies of texture zeros in neutrino physics~\cite{Fritzsch2011,Fukuda:1998mi,SK1998,Petcov2001,Grimus2004}.

\subsection{Texture $T_2$}

The texture $T_2$, featuring a vanishing $(2,2)$ element, provides a partial fit to the observed neutrino oscillation data. While the atmospheric mass-squared difference $\Delta m^2_{31}$ and two of the mixing angles are within the $3\sigma$ experimental ranges.

\[
M_\nu \approx \begin{pmatrix}
-0.02006 & -0.00319 + 0.00454i & -0.00883 + 0.00648i \\
-0.00319 - 0.00454i & 0 & 0.03037 + 0.01680i \\
-0.00883 - 0.00648i & 0.03037 - 0.01680i & 0.02616
\end{pmatrix}
\]

Texture that produced the next observables:

\[
\begin{aligned}
m_1 = -0.020929~\text{eV}, \,\,\,
m_2 = -0.025179~\text{eV}, \,\,\,
m_3 = \phantom{-}0.052208~\text{eV}, \,\,\,
\sum_{i=1}^3 |m_i| = 0.0983~\text{eV}.
\end{aligned}
\]

\[
\begin{aligned}
\Delta m_{21}^2 = 1.96 \times 10^{-4}~\text{eV}^2, \,\,\,
\Delta m_{31}^2 = 2.29 \times 10^{-3}~\text{eV}^2
\end{aligned}
\]

\[
U_{PMNS} = \begin{pmatrix}
0.840 & 0.517 & 0.166 \\
0.036 + 0.503i & 0.025 - 0.659i & -0.259 - 0.495i \\
-0.045 - 0.196i & 0.290 + 0.463i & -0.677 - 0.451i
\end{pmatrix}
\]

\[
\begin{aligned}
\theta_{12} = 31.61^\circ, \,\,\,
\theta_{13} = 9.53^\circ, \,\,\,
\theta_{23} = 34.46^\circ,\,\,\,
\end{aligned}
\delta = 223.4^\circ
\]

All of these values are evaluated with the best fit point for the minimum $\chi^2=0.39$. A vanishing (2,2) element, partially accommodates the current neutrino oscillation data. While $\Delta m^2_{31}$ and two mixing angles lie within the $3\sigma$ experimental ranges, the solar mass-squared difference $\Delta m^2_{21}$ and atmospheric angle $\theta_{23}$ are significantly outside the preferred bounds. Thus, although textures with zeros are theoretically motivated for reducing parameter space \cite{Frampton2002, Xing2002}, the $T_2$ structure is disfavored by present data \cite{JOUR,Alcaide2020,PDG}.

\subsection{Texture $T_3$}

The texture $T_3$, featuring an input zero in $(3,3)$, the form of the mass matrix that we obtain is: 

\[
M_\nu \approx \begin{pmatrix}
-0.03041 & -0.00996 + 0.00705i & -0.00871 - 0.00019i \\
-0.00996 - 0.00705i & 0.02277 & 0.03326 + 0.02773i \\
-0.00871 + 0.00019i & 0.03326 - 0.02773i & 0
\end{pmatrix}
\]
\\
Observables:

\[
\begin{aligned}
m_1 = -0.032442~\text{eV}, \,\,\,
m_2 = -0.033881~\text{eV}, \,\,\,
m_3 = \phantom{-}0.058679~\text{eV}, \,\,\,
\sum_{i=1}^3 |m_i| = 0.1250~\text{eV}.
\end{aligned}
\]

\[
\begin{aligned}
\Delta m_{21}^2 = 9.54 \times 10^{-5}~\text{eV}^2, \,\,\,
\Delta m_{31}^2 = 2.39 \times 10^{-3}~\text{eV}^2,
\end{aligned}
\]

\[
U_{PMNS} = \begin{pmatrix}
0.802 & -0.574 & -0.166 \\
-0.269 + 0.064i & -0.557 - 0.044i & 0.629 + 0.464i \\
0.436 - 0.301i & 0.436 - 0.410i & 0.600 - 0.035i
\end{pmatrix}
\]

\[
\begin{aligned}
\theta_{12} = 35.63^\circ, \,\,\,
\theta_{13} = 9.54^\circ, \,\,\,
\theta_{23} = 52.42^\circ,\,\,\,
\end{aligned}
\delta = 328.7^\circ.
\]

These values are evaluated with the best fit point for the minimum $\chi^2=2.10$.
This texture is close to experimental expectations, with mixing angles and mass-squared differences lying within the bounds $3\sigma$.

\subsection{Texture $T_4$}

We find a mass matrix with the following form (note that, for this particular case, all entries are real):

\[
M_\nu = \begin{pmatrix}
-0.00610 & 0 & 0.02191 \\
0 & 0.03603 & 0.02242 \\
0.02191 & 0.02242 & 0.02762
\end{pmatrix}~\text{eV}
\]

Observables

\[
\begin{aligned}
m_1 = -0.02236~\text{eV}, \,\,\,
m_2 = \phantom{-}0.02411~\text{eV}, \,\,\,
m_3 = \phantom{-}0.05581~\text{eV}, \,\,\,
\sum_{i=1}^3 |m_i| = 0.1023~\text{eV}.
\end{aligned}
\]

\[
\begin{aligned}
\Delta m_{21}^2 = 8.12 \times 10^{-5}~\text{eV}^2, \,\,\,
\Delta m_{31}^2 = 2.61 \times 10^{-3}~\text{eV}^2.
\end{aligned}
\]

\[
U_{PMNS} = \begin{pmatrix}
-0.8155 & 0.5594 & 0.1484 \\
-0.3027 & -0.6308 & 0.7145 \\
0.4933 & 0.5377 & 0.6837
\end{pmatrix}
\]

\[
\begin{aligned}
\theta_{12} = 34.45^\circ, \,\,\,
\theta_{13} = 8.53^\circ, \,\,\,
\theta_{23} = 46.26^\circ, \,\,\,
\end{aligned}
\delta = 0^\circ\,\,\, \text{or}\,\,\, \pi \quad \text{(Jarlskog invariant } J = 0)
\]

This texture provides an excellent fit to the experimental data. All mixing angles and mass-squared differences fall within the $3\sigma$ ranges, and the CP-violating phase corresponds to a vanishing Jarlskog invariant, consistent with the imposed matrix structure. These values are evaluated with the best fit point for the minimum $\chi^2=4.61\times10^{-8}$.

\subsection{Texture $T_5$}

The mass matrix is found to have the form:

\[
M_\nu = \begin{pmatrix}
-0.00028 & 0.00774 + 0.00782i & 0 \\
0.00774 - 0.00782i & 0.02442 & 0.02114 + 0.00508i \\
0 & 0.02114 - 0.00508i & 0.03048
\end{pmatrix}~\text{eV}
\]

Observables:
\[
\begin{aligned}
m_1 = -0.00681~\text{eV}, \,\,\,
m_2 = \phantom{-}0.01097~\text{eV}, \,\,\,
m_3 = \phantom{-}0.05046~\text{eV}, \,\,\,
\sum_{i=1}^3 |m_i| = 0.0682~\text{eV}.
\end{aligned}
\]

\[
\begin{aligned}
\Delta m_{21}^2 = 1.53 \times 10^{-4}~\text{eV}^2, \,\,\,
\Delta m_{31}^2 = 2.55 \times 10^{-3}~\text{eV}^2
\end{aligned}
\]

\[
U_{PMNS} = \begin{pmatrix}
0.8243 & 0.5472 & -0.1453 \\
-0.3440 + 0.3476i & 0.3933 - 0.3974i & -0.4709 + 0.4759i \\
0.1476 - 0.2439i & -0.3225 + 0.5329i & -0.3772 + 0.6232i
\end{pmatrix}
\]

\[
\begin{aligned}
\theta_{12} = 33.58^\circ, \,\,\,
\theta_{13} = 8.35^\circ, \,\,\,
\theta_{23} = 42.58^\circ, \,\,\,
\end{aligned}
\delta = 0^\circ\,\,\, \text{or}\,\,\, \pi. \quad 
\]

This texture successfully reproduces the observed neutrino oscillation parameters with all values within the $3\sigma$ experimental ranges, and an effectively vanishing CP-violating phase. 
However, the minimum $\chi^2=37.65$. Although the model contains the necessary parameters to describe the primary observables, it is missing one or more parameters required to capture smaller but statistically significant variations in the data. 
In this form, the number of free parameters (degrees of freedom) must be increased to see if the value of $\chi^2$ decreases to an acceptable level.

\subsection{Texture $T_6$}

We find a Hermitian neutrino mass matrix of the form:

\[
M_\nu = \begin{pmatrix}
0.05467 & 0.00173 + 0.00467i & 0.00111 + 0.00461i \\
0.00173 - 0.00467i & 0.02970 & 0 \\
0.00111 - 0.00461i & 0 & -0.00394
\end{pmatrix}~\text{eV}
\]

The observables obtained are: 
\[
\begin{aligned}
m_1 = 0.06352~\text{eV}, \,\,\,
m_2 = -0.06633~\text{eV}, \,\,\,
m_3 = -0.07548~\text{eV}, \,\,\,
\sum_{i=1}^3 |m_i| = 0.2053~\text{eV}.
\end{aligned}
\]

\[
\begin{aligned}
\Delta m_{21}^2 = 3.68 \times 10^{-4}~\text{eV}^2, \,\,\,
\Delta m_{31}^2 = 1.43 \times 10^{-3}~\text{eV}^2
\end{aligned}
\]

\[
U_{PMNS} = \begin{pmatrix}
0.9251 & 0.2856 & 0.2500 \\
-0.3713 + 0.0296i & 0.8077 - 0.0644i & 0.4511 - 0.0360i \\
0.0689 + 0.0250i & 0.4812 + 0.1744i & -0.8048 - 0.2917i
\end{pmatrix}
\]

\[
\begin{aligned}
\theta_{12} = 17.16^\circ, \,\,\,
\theta_{13} = 14.48^\circ, \,\,\,
\theta_{23} = 27.86^\circ, \,\,\,
\end{aligned}
\delta = 0^\circ\,\,\, \text{or}\,\,\, \pi. \quad 
\]

This texture fails to reproduce the experimental values of the mixing angles within the $3\sigma$ range and yields an unacceptably high sum of masses. It is therefore disfavored by current data.

\section{Inverted Ordering}
In the oscillations experiments have precisely measured the two mass-squared differences, $\Delta m_{21}^2$ and $|\Delta m_{21}^2|$, the ordering of the mass eigenstates remains an open quiestion. The inverted Hirarchy (IH) corresponds to the mass ordering $$m_3<m1<m2,$$ in which the two heavier mass eigenstate $m_1$ and $m_2$ form a close pair separated from the lighter state $m_3$.  This constrasts with teh normal hierarchy (NH), where $m_3$ is the heaviest.

\subsection{Numerical analysis}

The numerical analysis for the inverted hierarchy follows the same procedure described in Section~\ref{sec:4}. In particular, we employ the same texture structures introduced in Section~\ref{sec:4.1}. Among the considered textures, only the T2 texture successfully reproduces the experimental data within the $1\sigma$ level. The methodology used to obtain these results is outlined below.

\noindent  The T2 texture is defined as
\[
T_2 =
\begin{pmatrix}
a & b & c \\
b^{*} & 0 & d \\
c^{*} & d^{*} & e
\end{pmatrix}.
\]

\noindent The diagonalization procedure was carried out under the constraint $m_2 > m_1 > m_3$, appropriate for the inverted mass hierarchy. This approach allows us to express the UPMNS matrix in terms of the parameters $m_1,\ m_2,\ m_3$ and $b$.\\
Applying the $\chi^2$ minimization method described in Section~\ref{sec:4.2} yields the numerical results presented below.

\[
\begin{aligned}
m_1 = 0.0561248~\text{eV}, \,\,\,
m_2 = 0.0567404~\text{eV}, \,\,\,
m_3 = 0.0256819~\text{eV}, \,\,\,
\sum_{i=1}^3 |m_i| = 0.138~\text{eV}.
\end{aligned}
\]
The calculated $\chi^2$ statistic was $2.21\times 10^{-8}$ with three degrees of freedom ($df = 3$)

\[
\begin{aligned}
\Delta m_{21}^2 = 6.95 \times 10^{-5}~\text{eV}^2, \,\,\,
\Delta m_{31}^2 = 2.50 \times 10^{-3}~\text{eV}^2.
\end{aligned}
\]

\[
U_{PMNS} = \begin{pmatrix}
0.821372 & 0.550543 & 0.149164 \\
-0.47604 & 0.517586 & 0.710979 \\
0.314219 & -0.654986 & 0.687211
\end{pmatrix}
\]

\[
\begin{aligned}
\theta_{12} = 34.54^\circ, \,\,\,
\theta_{13} = 8.54^\circ, \,\,\,
\theta_{23} = 46.13^\circ, \,\,\,
\end{aligned}
\delta = 0^\circ\,\,\, \text{or}\,\,\, \pi \quad \text{(Jarlskog invariant } J = 0)
\]

\section{Conclusions}

We have systematically analyzed the six possible one zero texture patterns for Hermitian neutrino mass matrices under the assumption of both neutrino normal and inverted mass hierarchy, working in the charged lepton diagonal basis. In this scenario, the lepton mixing matrix arises entirely from the neutrino sector. Mathematically, since the number of free parameters exceeds the number of physical observables, analytical solutions always exist. Specifically, the mass-squared differences, the three mixing angles, and the CP-violating phase can all be determined analytically for each texture. A crucial observation is that when the texture zero appears off the diagonal, the Jarlskog invariant vanishes, implying that the Dirac CP-violating phase must be either $\delta = 0 $ or $\delta = \pi$. Therefore, nontrivial CP violation $( \delta \neq 0, \pi )$  is only possible when the texture zero is located in one of the diagonal entries of the mass matrix.\\

Our numerical analyses in the normal ordering reveal that:

\begin{itemize}
    \item \textbf{Texture $T_1$} yields an excellent fit to all current experimental data. All three mixing angles, the mass-squared differences, and the CP phase are reproduced within 1$\sigma$ of the deviation of the global best fit values. In particular, the best fit CP phase is found to be close to  $\delta \approx 270^\circ $, and the total sum of neutrino masses is consistent with cosmological bounds, such as those from supernova neutrino constraints and the Planck satellite.

    \item \textbf{Texture  $T_2$} fails to accommodate the experimental data, with significant deviations in the atmospheric mixing angle and solar mass-squared difference.

    \item \textbf{Texture $T_3$} provides aviable fit to all observables within 3$\sigma$.

    \item In the remaining textures $T_4$,  $T_5$, and $T_6$ the phase of CP violation is necessarily trivial $( \delta = 0 $  or  $\pi$), as expected due to the texture zero being located off-diagonal. However,
    \begin{itemize}
        \item Texture $T_4$, which is real, and
        \item Texture  $T_5$, which is complex,
    \end{itemize}
    both successfully reproduce all observables within 3$\sigma$.

    \item \textbf{Texture \( T_6 \)} fails to match current experimental constraints and is thus disfavored.
\end{itemize}

Our numerical analyses in the inverted ordering reveal that:

\begin{itemize}
    \item \textbf{Texture $T_2$} fit to current experimental information is excellent. All key parameters—the three mixing angles, the mass differences, and the CP phase—agree with the global best-fit values at the $1\sigma$ level of deviation. The optimal CP phase is found to be near $270^\circ$, and the total neutrino mass is in conformance with cosmological bounds, including those from supernovae and the Planck satellite.
    \item The other five single texture zero configurations are found to be in disagreement with the current experimental data. Therefore, they must be ruled out (or excluded) as viable solutions within this framework.
\end{itemize}

In summary, we have shown that although analytic solutions always exist for Hermitian neutrino mass matrices with a single texture zero, not all such patterns yield physically acceptable predictions. \\

\section*{Acknowledges}
 R. H. B. acknowledges additional financial support from Minciencias CD82315 CT ICETEX 2021-1080.

\bibliographystyle{unsrt}
\bibliography{referencias}

\end{document}